\documentclass[aps,superscriptaddress,nofootinbib,showpacs]{revtex4}\usepackage{amssymb}
\usepackage{amsmath}
\usepackage{amsthm}
\usepackage{graphicx}
\usepackage{graphics}

\newcommand\scri{\mathcal{I}}

\newtheorem{thm}{Theorem}

\newtheorem{defn}{Definition}
\theoremstyle{remark}

\newtheorem{exam}{Example}

\begin{document}

\title{What does Birkhoff's theorem really tell us?}
\author{Kristin Schleich}
\author{Donald M. Witt}
\affiliation{Department of Physics and Astronomy, University of British Columbia,
Vancouver, British Columbia \ V6T 1Z1}
\affiliation{Perimeter Institute for Theoretical Physics, 31 Caroline Street North, Waterloo,
ON, N2L 2Y5, Canada}
\date{\today}

\begin{abstract}
  Birkhoff's theorem is a classic result that characterizes locally spherically symmetric solutions of the Einstein equations.
In this paper, we illustrate the consequences of its local nature  for the cases of vacuum and  positive cosmological constant.   We construct several examples of initial data for spherically symmetric spacetimes on Cauchy surfaces of different  topology than ${\mathbb R}\times S^2$,  that of the maximal analytic extension of Schwarzschild and Schwarzschild-de Sitter spacetimes. The spacetimes formed from the evolution of these initial data sets also have very different physical properties; in particular they need not contain a static region or be asymptotically flat or asymptotically de Sitter. We also present locally spherically symmetric initial data sets for de Sitter spacetimes that are not covered by the maximal analytic extension of de Sitter spacetime itself.  Finally we illustrate the utility of Birkhoff's theorem in identifying the spacetimes associated with two spherically symmetric initial data sets; one proven to exist but not explicitly exhibited and  one which has negative ADM mass.
\end{abstract}
\maketitle

\section{Introduction}

Birkhoff's theorem \cite{Birkhoff,Jebsen,Alexandrow,Eiesland1925}, a result that characterizes the structure of  locally spherically symmetric vacuum spacetimes,  is not only a classic contribution to general relativity but is also an important tool in gravitational physics and cosmology.
 Considered general relativity's analogue of Newton's iron 
sphere theorem (see, for example, \cite{pe}), it  is invoked both in mathematical relativity and in more general contexts. For example, it is used to argue that empty spacetime interior to that of a spherically symmetric mass shell is Minkowski spacetime. It is also used to show that spherical gravitational collapse 
does not radiate gravitational waves.  Furthermore, it is used implicitly to draw the final conclusion in the proof of the classic uniqueness theorem for static black holes \cite{Israel:1967wq}. Hence, Birkhoff's theorem has played an essential role in the study of spherically symmetric spacetimes in general relativity.

Birkhoff's theorem shows that any spherically symmetric solution of the vacuum Einstein equations is
locally isometric to a neighborhood in Schwarzschild spacetime. Hence it is a local uniqueness theorem whose corollary is that locally spherically symmetric solutions  exhibit an additional local Killing vector field; however this field is not necessarily timelike.  The local nature of Birkhoff's theorem is clear in various classic proofs and in many proofs of its generalizations to other matter sources in gravitational physics such as cosmological constant. However, these proofs usually stop at the result; hence the consequences of its local nature may not be apparent. 

The purpose of this paper is to remedy this gap by providing examples illustrating the local nature of Birkhoff's theorem. We construct initial data sets for spacetimes that satisfy the conditions for Birkhoff's theorem that have global structure different than that of the maximal analytic extension of Schwarzschild. These examples do not  always exhibit  other properties of the maximal analytic extension;  in particular they are not all asymptotically flat. Furthermore, some do not exhibit a static region.  We then provide similar initial data sets for spacetimes that satisfy Birkhoff's theorem generalized to positive cosmological constant and note an analogous variance of their properties from those of the maximal analytic extension of Schwarzschild-de Sitter spacetime. Finally we note that the local nature of Birkhoff's theorem provides a simple, useful tool for the identification of spacetimes resulting from the evolution of spherically symmetric data sets.  Therefore, recognition that Birkhoff's theorem is still applicable to spherically symmetric spacetimes of more general topology is useful in the analysis of the their physical properties.

In Section 2, we prove a needed theorem that connects spherically symmetric initial data sets to spherically symmetric spacetimes. In Section 3 we construct global spherically symmetric initial data sets which evolve to spacetimes  whose topology and various physical properties are not those of the maximal analytic extensions of Schwarzschild or Schwarzschild-de Sitter spacetime. We first give examples for the vacuum case, then extend these results to the case of positive cosmological constant. In Section 4, we construct examples of locally spherically symmetric initial data sets, first by identification on Minkowski and de Sitter ones, then by a more general procedure for the de Sitter case. These more general de Sitter initial data sets have generic topology and it has been shown that, generically, their universal cover is not de Sitter spacetime itself \cite{ MorrowJones:1988yw,MorrowJones:1993zu}. In Section 5, we demonstrate that Birkhoff's theorem can be used to identify the spacetimes associated with an abstract initial data set and with an explicitly constructed initial data set that has negative ADM mass. 

Birkhoff's theorem for the case of cosmological constant demonstrates that the local geometry of a spherically symmetric spacetime is that of a neighborhood in
Schwarzschild(-anti-de Sitter) spacetime for zero (negative) cosmological constant,  and in either Schwarzschild-de Sitter or Nariai spacetime  for positive cosmological constant. Various proofs of Birkhoff's theorem for cosmological constant  are found in \cite{Eiesland1925,MorrowJones:1988yw,MorrowJones:1993zu,rindler,Schleich:2009uj}. That provided in \cite{ MorrowJones:1988yw,MorrowJones:1993zu}  explicitly exhibits both the Schwarzschild-de Sitter and Nariai solutions; this is also done in that of \cite{rindler} without citation of previous references. A simple unified proof was provided in \cite{Schleich:2009uj}. This paper follows its notation.

\section{Initial Data for Spherically Symmetric Spacetimes}
\label{section3}
Generalized Birkhoff's theorems are local uniqueness theorems  that fix the  local geometry of a spherically symmetric spacetime. In order to discuss the possible global structures of such spacetimes, it is useful to tie Birkhoff's theorem to a formulation in which the global structure is readily apparent. One way to do so is to construct locally spherically symmetric spacetimes from initial data sets as this formulation allows the specification of the global topology in a natural way, through that of the Cauchy surface. This formulation is also of interest due to its use in numerical relativity. To achieve the tie between initial data and Birkhoff's theorem, one needs to show that spherically symmetric initial data evolves to form spherically symmetric spacetimes. We do so below.

First recall that the Einstein equations for a globally hyperbolic spacetime  $\Sigma \times {\mathbb R}$ with
4-metric $g_{ab}$ can be written in initial value form on the Cauchy slice $\Sigma $ by introducing
coordinates such that $\Sigma$ is a constant time slice.
The normal vector to $\Sigma$ is  $$n^a = \frac {1}{N}(t^a - N^a)$$ where $t^a$ is the
tangent vector to the time function $t$. $N$ is usually termed the lapse and $N^a$ the shift. Then
$$ g_{ab} = -n_a n_b + h_{ab}$$
where $h_{ab}$ is a geodesically complete
Riemannian metric on $\Sigma$. The extrinsic curvature of $\Sigma$  is given by
\begin{equation}K_{ab}=\frac 12 {\mathcal L}_n{h_{ab}}\label{Kdef}\end{equation}
 where $ {\mathcal L}_n$ is the Lie derivative with respect to vector $n$.
The 
3-metric $h_{ab}$ is the restriction of the spacetime metric $ g_{ab}$  to $\Sigma$ and extrinsic curvature $K_{ab}$ is the 
rate of change of the 3-metric in the spacetime. 
The 3-metric and extrinsic 
curvature satisfy the constraints:
\begin{align}R-K_{ab}K^{ab}+K^2&=16\pi \rho+ 2\Lambda\cr
D_b(K^{ab}-Kh^{ab})&=8\pi J^a\label{constraints}\end{align}
where $R$ is the scalar curvature of $h_{ab}$ and $D_a$ the compatible covariant derivative, $\rho$ and $J^a$
are the energy and momentum densities of the matter sources respectively and $\Lambda$ the cosmological constant. For vacuum initial data, $\rho$ and $J^a$ both vanish.  The time evolution of $h$ and $K$ is given through (\ref{Kdef}) and the remaining  Einstein equations,
\begin{equation}R^a_b - \frac {1}{2}R h^a_b + {\mathcal L}_{n} (K^a_b-h^a_bK) -KK^a_b+ 
 \frac {1}{2} h^a_bK^2-\frac {1}{2}h^a_bK^{cd}K_{cd}= -\Lambda \, h^a_b \ ,\end{equation}
 for the case of vacuum initial data with cosmological constant.

Conversely, given any 3-manifold $\Sigma$ with spatially geodesically complete Riemannian metric $h_{ab}$ and symmetric tensor 
$K_{ab}$ satisfying the constraints (\ref{constraints}) with physically reasonable matter sources (precisely, matter sources that satisfy the dominant energy condition, $\rho+\frac{\Lambda}{8\pi} \geq (J^aJ_a)^{\frac 12}$), one can the evolve the initial data $h_{ab}$ and $K_{ab}$ into a globally hyperbolic spacetime of topology  $\Sigma \times {\mathbb R}$ \cite{he}. 

The definition of a locally spherically symmetric space, Definition 3 of \cite{Schleich:2009uj},  has a natural extension  to initial data. 
\begin{defn} \label{defninit} Initial data is locally
spherically symmetric if both $h_{ab}$ and $K_{ab}$ are locally
spherically symmetric tensors on the Cauchy slice $\Sigma $.\end{defn}

Next we prove that locally spherically symmetric initial data evolves to form a locally spherically symmetric spacetime:
\begin{thm} \label{init} Let $( h_{ab}, K_{ab} )$ be locally spherically symmetric initial data on
$\Sigma$ which satisfies the vacuum constraints (\ref{constraints})  with nonnegative cosmological constant.
Then this initial data evolves into a spacetime $M^4$ such that, for some open 
neighborhood of $\Sigma$ in $M^4$, the spacetime metric is also locally spherically symmetric.
\end{thm}

\begin{proof}
As the initial data is locally spherically symmetric,  there is a set of three independent vector fields $\{\xi _i{}\}$, $i= 1, 2,3$ in an open neighborhood $U_p$ around any point $p$ in $\Sigma$ that generate the Lie algebra of $SO(3)$.
The {\it synchronous gauge}, $N=1$ and $N^a=0$, is  always a 
valid local gauge choice, that is in a sufficiently small neighborhood of $p$ in $\Sigma$, for any spacetime satisfying the constraints. Observe that  $t^a$ is equal to the normal to the hypersurface for this choice. Using this gauge,
the spacetime metric in an open neighborhood $U$ of $M^4$ whose intersection with $\Sigma$ is 
the open neighborhood $U_p$ can be written as 
\begin{equation*}ds^2=-dt^2+h_{\alpha\beta}(t)dx^\alpha dx^\beta\end{equation*}
where the coordinates $x^\alpha$ are spatial coordinates on $\Sigma$.  
The vector fields $\{\xi _i{}\}$, $i= 1, 2,3$ on $U_p$  can be 
trivially extended to $U$ by requiring that each have no normal component and 
that they be time independent;  $\xi  _i(x,t)=\xi_i(x)$  and $t\cdot \xi_i(x,t) = 0$ in $U$.
Each of these local Killing vector fields $\xi _i (x,t)$ generates a corresponding flow
$ \Psi _i$ on $U$. 

Let $(h_{ab}(t), K_{ab}(t))$ represent the time evolution of the initial
data $(h_{ab}, K_{ab})$ with time chosen so that $h_{ab}(0)=h_{ab}$ and $K_{ab}(0)=K_{ab}$. As the Cauchy problem is well posed for nonnegative cosmological constant \cite{he}, this evolution exists and is unique. 
Now, in the neighborhood $U$,  $ (\Psi_i h_{ab}(t), \Psi_iK_{ab}(t))$ is also a solution to the 
Einstein equations with initial data $( \Psi_ih_{ab}, \Psi_iK_{ab})$ in the corresponding neighborhood $U_p$ in $\Sigma$
by diffeomorphism invariance. However, $( \Psi_ih_{ab}, \Psi_iK_{ab})=( h_{ab}, K_{ab} )$ as the initial data is locally spherically symmetric. The cosmological constant is also invariant under $\Psi_i$.
Thus  $ (\Psi_i h_{ab}(t), \Psi_iK_{ab}(t))$ is a solution to the Einstein equations in $U$  for the same initial data as $(h_{ab}(t), K_{ab}(t))$. 
It follows, by the uniqueness of the evolution of the initial data, that $ \Psi  _ih_{ab}(t)=h_{ab}(t)$ and $ \Psi  _iK_{ab}(t)=K_{ab}(t)$ in $U$.
Furthermore, the orbits of $SO(3)$ do not change dimension or become spacelike under local time evolution. 
Therefore, the spacetime is locally spherically symmetric. 
\end{proof}

This theorem can clearly be generalized to  global spherical symmetry. In addition, it can be extended to the case of negative cosmological constant, a case that violates the dominant energy condition as it has negative energy density,  by utilizing an alternate proof of the existence and uniqueness of a solution to the Einstein equations for nonzero cosmological constant, for example, that of \cite{dt}. 

Note that the above proof is abstract; no coordinate choice, i.e.  gauge choice, has been used in the specification of the initial data $(h_{ab},K_{ab})$. Instead, the coordinate choice is made in the specification of synchronous gauge. If the initial data is specified explicitly in a coordinate chart on $\Sigma$, this freedom to freely specify synchronous gauge is removed. Instead, one must solve for $N$ and $N^a$. Of course, it  is often easiest to present initial data explicitly on a 3-manifold by introducing a coordinate chart, for example, one that explicitly exhibits spherical symmetry. Thus the examples in the next Section, in general, do not have $N = 1$ due to the explicit choice of coordinates on the 3-manifold. Even so, Theorem \ref{init} still guarantees that these examples are locally spherically symmetric spacetimes and consequently, Birkhoff's theorem applies.

\section{Globally spherically symmetric initial data sets}\label{section4}
Armed with the results of the previous Section, we now proceed to construct examples of initial data sets for spacetimes that satisfy Birkhoff's theorem but are not the maximal analytic extensions discussed in \cite{Schleich:2009uj}. We first do so for the vacuum case in  \ref{section4a} then generalize these examples to the case of positive cosmological constant in  \ref{section5}.

\subsection{The vacuum case}\label{section4a}
We now construct spherically symmetric vacuum initial data sets on Cauchy surfaces of several different topologies. Theorem \ref{init} implies that their evolution results in a spacetime that satisfies the conditions needed for Birkhoff's theorem; hence these initial data sets are those for locally Schwarzschild spacetimes. However, the choice of nontrivial topology results in spacetimes that are not the maximal analytic extension of Schwarzschild spacetime. Consequently, these spacetimes do not exhibit various characteristics of this particular solution.  We begin with two simple initial data sets whose Cauchy surface has different topology than ${\mathbb R}\times S^2$, that of the maximal analytic extension of Schwarzchild. The first has time symmetric initial data:
\begin{exam}(${\mathbb R}P^2$ Schwarzschild) \label{rp2}
Take the Cauchy surface to have topology
$\Sigma ={\mathbb R}\times {\mathbb R}P^2$. Introducing the notation ${\bf h}= h_{\alpha\beta}dx^\alpha dx^\beta$,  take the metric to be
\begin{equation}\label{sinit}{\bf  h }=  (1+\frac {z^2}{16M^2}) dz^2 + 4M^2(1+\frac {z^2}{16M^2})^2 d\Omega_2^2\\
\end{equation}
 where  $d\Omega_2^2$ is the round metric on ${\mathbb R}P^2$.
Strictly speaking, this metric is given above in a chart $U = {\mathbb R}\times U_{{\mathbb R}P^2}$, where $U_{{\mathbb R}P^2}$ is a neighborhood on  $ {\mathbb R}P^2$ but it can be extended to  a set of charts covering all of
${\mathbb R}\times{\mathbb R}P^2$ in the obvious way.
This metric is spatially geodesically complete and is manifestly globally spherically symmetric.
A change of coordinates, $z= \sqrt{8Mr-16M^2}$, yields
\begin{equation}
{\bf h}=  \frac {dr^2}{ 1-\frac {2M}{r}}+ r^2d\Omega _2^2\ , \label{usualS} \end{equation}
the spatial metric on the time symmetric slice of Schwarzschild spacetime in Schwarzschild coordinates. 
This form of the metric is, of course, in a chart for which $r>2M$; the limit  $r\to 2M$ corresponds to $z\to 0$ in (\ref{sinit}) and thus $r= 2M$ is clearly a standard coordinate singularity. 
  
 The choice $K_{ab}=0$ with metric (\ref{sinit}) satisfies the constraints (\ref{constraints}) as $R(h) = 0$. This initial data is manifestly globally spherically symmetric; hence its maximal evolution is also globally spherically symmetric.  
The resulting spacetime  is  the maximal analytic extension of 
Schwarzschild spacetime with the identification ${\mathbb Z}_2=\{ I, P\}$ on 2-spheres in the appropriate set of spatial hypersurfaces.  This spacetime is clearly only locally Schwarzschild as its topology is ${\mathbb R}^2 \times {\mathbb R}P^2$.  It is  also not 
asymptotically flat  as $\scri$ has topology ${\mathbb R} \times {\mathbb R}P^2$. 

\end{exam}

\begin{exam}( The ${\mathbb R}P^3$ geon) \label{rp3}  An alternate construction of this spacetime was given  in \cite{Friedman:1993ty}. 
Take the Cauchy surface to have the topology of  a punctured  ${\mathbb R}P^3$,  ${\mathbb R}P^3-\{p\}$. 
This manifold admits a globally spherically symmetric metric: Observe that  ${\mathbb R}P^3$ can be constructed by identifying antipodal points on $S^3$. Explicitly,
the  3-sphere with its round metric can be realized as $S^3=\{(x,y,z,w)\in { \mathbb R}^4| x^2 + y^2+z^2+w^2=1\}$. Next take ${\mathbb Z}_2$ to be  generated by the map
${\hat P}$ on $v\in {\mathbb R}^4$ given by $v\rightarrow -v$ restricted to the 3-sphere. Then ${\mathbb R}P^3=S^3/{\mathbb Z}_2$. In contrast to the similar construction of ${\mathbb R}P^2$, this map is orientation preserving; thus ${\mathbb R}P^3$
is orientable. Next, observe that the universal cover of  the  ${\mathbb R}P^3 - \{p\}$ is 
${\mathbb R}\times S^2$ which is topologically the same as double punctured $S^3$, that is $S^3$ with the ``north'' and ``south'' poles  removed.  Note that ${\mathbb Z}_2=\{ I, {\hat P}\}$  acts freely on this space.  Thus ${\mathbb R}\times S^2$ allows geometries that are invariant under both  $SO(3)$ and ${\mathbb Z}_2$. It follows that ${\mathbb R}P^3 - \{p\}$ also admits an $SO(3)$ invariant geometry. This intuitive argument can be made rigourous;   a demonstration that the isometry group  of ${\mathbb R}P^3 - \{p\}$ includes $SO(3)$ is presented in the Appendix.

Time symmetric initial data on the cover of ${\mathbb R}P^3 - \{p\}$, ${\mathbb R}\times S^2$,  is given by metric (\ref{sinit}) with $K_{ab}=0$. The map ${\mathbb Z}_2=\{I,\hat P\}$ acts freely on this initial data set; in particular it identifies points with $z<0$ to  antipodally related points with $z>0$. At $z=0$ this map reduces to an antipodal map on the minimal $S^2$; it produces an ${\mathbb R}P^2$. This also realizes an alternate construction of ${\mathbb R}P^3 - \{p\}$; the attachment of  ${\mathbb R}\times S^2$ to ${\mathbb R}P^2$. This will be useful in other examples. As this sphere at $z=0$ is minimal, the identification is totally geodesic; consequently, the metric $\bar h$ induced by this identification is smooth.  Consequently,  $\bar h$ with $K_{ab}=0$  is a smooth initial data set on ${\mathbb R}P^3 - \{p\}$.

One can find the evolution of this initial data for the  ${\mathbb R}P^3$ geon explicitly. The group ${\mathbb Z}_2=\{ I, {\hat P}\}$ also acts 
freely on the evolution of the Schwarzschild initial data on the covering space, that is Schwarzschild spacetime. In Kruskal coordinates, the Schwarzschild spacetime can be written as
\begin{equation}
ds^2 = \frac {32 M^3}{r} e^{-\frac r{2M}}( -dT^2 + dX^2) + r^2 d\Omega_2^2\label{Kruskal}\end{equation}
where $r$ is defined implicitly by $T^2 - X^2 = (1-\frac {r}{2M})e^{\frac r{2M}}$ where $r$ is that of (\ref{usualS}). Note that in the coordinate $z= \sqrt{8Mr-16M^2}$, this relation is $T^2-X^2 = \frac{z^2}{16M^2}e^{(1+ \frac{z^2}{16M^2})}$. It follows that the Kruskal metric on the $T=0$ slice is simply (\ref{sinit}). Therefore the identification on Schwarzschild spacetime given by  $\hat P: X \to -X$ generates the evolution of the initial data set for the  ${\mathbb R}P^3$ geon. The Penrose diagram of this spacetime is given in Figure \ref{fig:rp3}. 
 \begin{figure}
\includegraphics{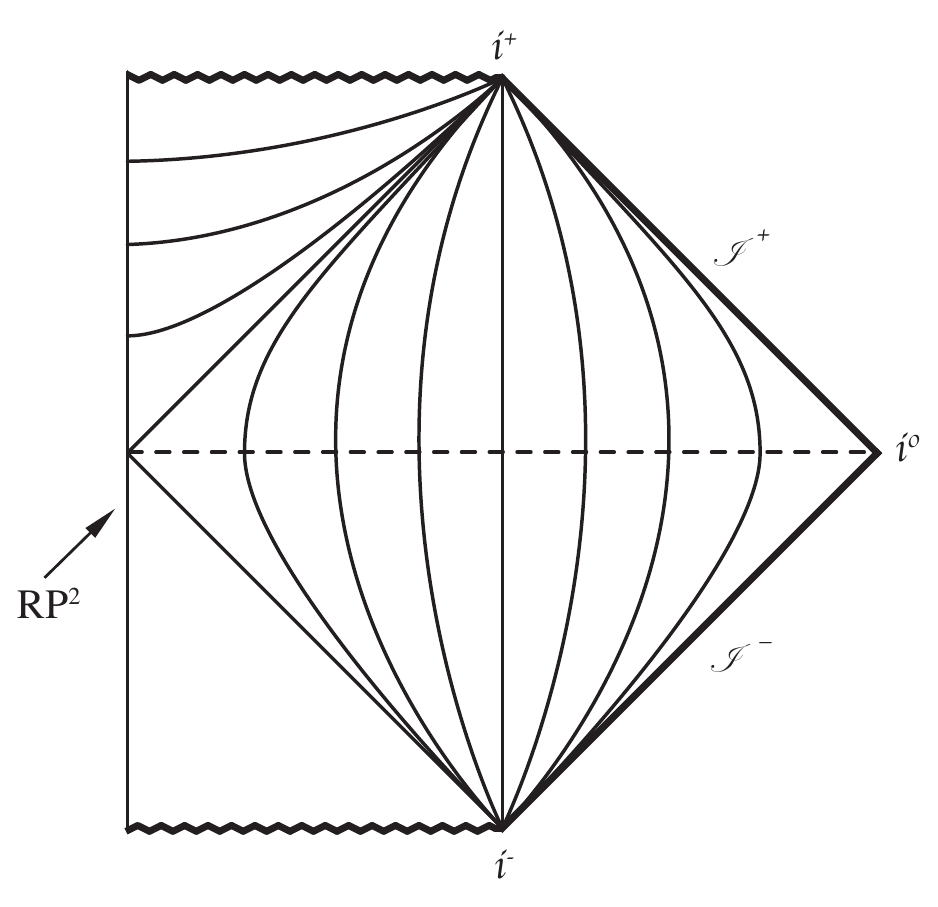}
\caption{The Penrose diagram for the ${\mathbb R}P^3$ geon.  Integral curves of the Killing vector field  have been sketched in the exterior and black hole regions. The dotted line is the zero extrinsic curvature Cauchy surface for this spacetime. Note that its intersection with the horizon is ${\mathbb R}P^2$.}
\label{fig:rp3}
\end{figure}

The ${\mathbb R}P^3$ geon  has a single asymptotically flat region and its ADM mass can take any positive value. However,  this time symmetric initial data set for the ${\mathbb R}P^3$ geon does not contain a trapped surface, in contrast to the situation in covering space, the Schwarzschild initial data set.  The minimal surface ${\mathbb R}P^2\subset {\mathbb R}P^3 - \{p\}$ in the initial data
non-orientable. However, ${\mathbb R}P^3 - \{p\}$ is orientable. Therefore the minimal ${\mathbb R}P^2$ must have a non-trivial normal bundle; in 
other words, it is only one sided in ${\mathbb R}P^3 - \{p\}$. 
This means that although ${\mathbb R}P^2$ is a marginally trapped set in this initial data set, it  is not  a trapped surface because a trapped surface must be two sided; it
must separate the Cauchy surface into an inside region and  outside region.\footnote{Note that as the Cauchy slice for ${\mathbb R}P^2$ Schwarzschild,
$\Sigma ={\mathbb R}\times {\mathbb R}P^2$,
is non-orientable, the minimal  ${\mathbb R}P^2$ surface at $r=0$  is  two sided  and thus is a trapped surface. } It follows that, although a spacelike cut of the black hole horizon in the evolution of this initial data at a time future to this surface has topology $S^2$, the horizon generators begin on ${\mathbb R}P^2$. 
This interesting property of the ${\mathbb R}P^3$ geon is an illustration of subtleties in the determination of the topology of cuts of the horizon by spacelike hypersurfaces \cite{Friedman:1993ty,Galloway:1999bp}.

\end{exam}

The next example has nonzero extrinsic curvature.

\begin{exam}(Schwarzschild Cosmologies) \label{sc}

Take  the Cauchy surface to have topology ${\mathbb R}\times S^2$ and choose the metric 
\begin{equation}\label{hdata}{\bf  h}=a^2 \left(d\psi^2 +d\Omega ^2\right)\end{equation}
where $a$ is a constant. The scalar curvature of $h_{ab}$ is simply $2/a^2$.  Introducing the notation ${\bf K} = K_{ab}dx^adx^b$, choose
 \begin{equation}\label{kdata}{\bf K}=b d\psi^2 + c d\Omega ^2\end{equation}
where  $b$ and $c $ are also constants. This choice of initial data trivially satisfies the momentum constraint.  The hamiltonian constraint yields the relation
\begin{equation}\label{sconstraint}
\frac 1{a^2} + \frac {2bc}{a^4} + \frac {c^2}{a^4} = 0 
\end{equation}
between the  three constants. The evolution of this initial data set yields the Schwarzschild solution in the region interior to the black hole or white hole horizon,
\begin{equation}
ds^2 = -\frac{dt^2}{\left(  \frac {2M}{t} - 1\right) } + \left( \frac {2M}{t} - 1\right)dr^2 + t^2d\Omega _2^2 \ , \label{ssmetricinside}
\end{equation}
where 
\begin{equation}
M=-\frac {bc}{a} 
\end{equation}
 and $ dr= \alpha d\psi$ where  $\alpha$  is chosen
so that $(2M/a - 1)\alpha^2 = a^2$.
Note that $M$ is positive as $bc<0$ by (\ref{sconstraint}).  The sign of the extrinsic curvature distinguishes between the black and white hole cases. The Cauchy surface for this initial data set corresponds to the integral curves of the Killing vector field in either the black hole or white hole region of the maximal extension of Schwarzschild spacetime. (See, for example, Figure 1 of \cite{Schleich:2009uj}). The horizon is the Cauchy horizon for the evolution of this initial data; it lies either to the past of future of the surface, again as determined by the sign of the extrinsic curvature. However, this spacetime clearly can be extended through the horizon by a coordinate transformation, for example to Kruskal coordinates (\ref{Kruskal}).\footnote{ Explicitly, the $t,r$ coordinates of metric (\ref{ssmetricinside}) can be related to those of (\ref{Kruskal}) by $T = (1-\frac {t}{2M})^{\frac 12} e^{\frac t{4M}}\sinh (\frac r{4M})$ and $X =  (1-\frac {t}{2M})^{\frac 12} e^{\frac t{4M}}\cosh(\frac r{4M})$ in the black hole interior.}

This extension is no longer possible if one poses the initial data on a closed global topology derived by identifications. As both ${\bf h}$ and ${\bf K}$ are invariant under translations, every 2-sphere in the initial data is totally geodesic. Therefore one can construct closed topologies by identifying these spheres with an appropriate group action. These initial data sets yield closed cosmologies with nontrivial topology that are singular to both the past and future of the Cauchy surface.

 For example, $S^1\times S^2 = {\mathbb R}\times S^2/{\mathbb Z}$, where the identification map is  $\psi \to \psi + \kappa$  where $\kappa$ is a positive constant. The identification map with $\kappa=\beta/\alpha$ extends to $r \to r + \beta$ in the resulting evolution (\ref{ssmetricinside}) with $\Lambda=0$. In Kruskal coordinates, this map is
$T\to T\cosh \beta + X \sinh \beta$, $X\to X \cosh\beta + T\sinh \beta$. On the Cauchy (black  or white hole) horizon, $T=\pm X$, this map reduces to a dilatation, $T\to e^{\pm \beta} T$; consequently, the point $(0,0)$ is a fixed point of this map. Hence, the locally Schwarzschild spacetime with topology $S^1\times S^2\times {\mathbb R}$ cannot be extended through this surface. Therefore the $S^1\times S^2$ Schwarzschild spacetime is singular both to the future and the past; it is a cosmology with closed spatial hypersurfaces. 
 
Three other closed cosmologies with different topology can be similarly constructed. The non-orientable
handle $S^1{\tilde { \times }} S^2$ is constructed by identifying each point $p$ on the 2-sphere at $\psi$ to the antipodal point $Pp$ on that at $\psi + \beta/\alpha$.   A similar procedure after replacement of $S^2$ with  ${\mathbb R}P^2$ results in $S^1\times {\mathbb R}P^2$.\footnote{The identification yielding the open Cauchy surface ${\mathbb R}\times {\mathbb R}P^2$
can, of course, be extended; the resulting space is ${\mathbb R}\times {\mathbb R}P^2$ Schwarzschild.} 
Initial data with topology ${\mathbb R}P^3\#{\mathbb R}P^3 $ can also be constructed. First note that initial data with topology ${\mathbb R}P^3 - \{p\}$ can be formed from (\ref{hdata}) and (\ref{kdata}) by carrying out the antipodal identification on ${\mathbb R}\times S^2$ as in Example \ref{rp3}; its evolution is simply the interior solution of the
${\mathbb R}P^3$ geon.   The Cauchy surface ${\mathbb R}P^3\#{\mathbb R}P^3 $ is formed by a further identification of antipodal points on any $S^2$ a finite distance from that of the minimal ${\mathbb R}P^2$ in this initial data set on ${\mathbb R}P^3 - \{p\}$. As for the $S^1\times S^2$ Schwarzschild cosmology, the spacetimes arising from the initial data sets on the three closed manifolds $S^1\times {\mathbb R}P^2$, $S^1{\tilde { \times }} S^2$ and
 ${\mathbb R}P^3\#{\mathbb R}P^3 $ cannot be extended across the Cauchy horizon. Hence they are also singular to both the future and the past.

These solutions are clearly not asymptotically flat. Notably, they are also not locally static and cannot be extended to exhibit a locally static region; they have only a local translational Killing vector that is manifestly spacelike everywhere. 
\end{exam}

\subsection{The positive cosmological constant case}\label{section5}

The three  examples of Section  \ref{section5}  have natural generalizations to the case of positive cosmological constant.
Examples \ref{rp2} and \ref{rp3} both extend to the case of Schwarzschild-de Sitter spacetime with $0<\Lambda<\frac 1{9M^2}$, but exhibit additional complexity due to the structure of
Schwarzschild-de Sitter itself. The maximal analytic extension of Schwarzschild-de Sitter  with $0<\Lambda<\frac 1{9M^2}$ has an infinite number of asymptotically de Sitter and black hole regions (See, for example, Figure 3 a) in \cite{Schleich:2009uj}). This property allows for more diversity in the construction of locally Schwarzschild-de Sitter initial data sets than seen in the Schwarzschild case.
\begin{exam}(Topologically nontrivial Schwarzschild-de Sitter spacetimes for  $0<\Lambda<\frac 1{9M^2}$) \label{rpdesitter}

 \begin{figure}
\includegraphics{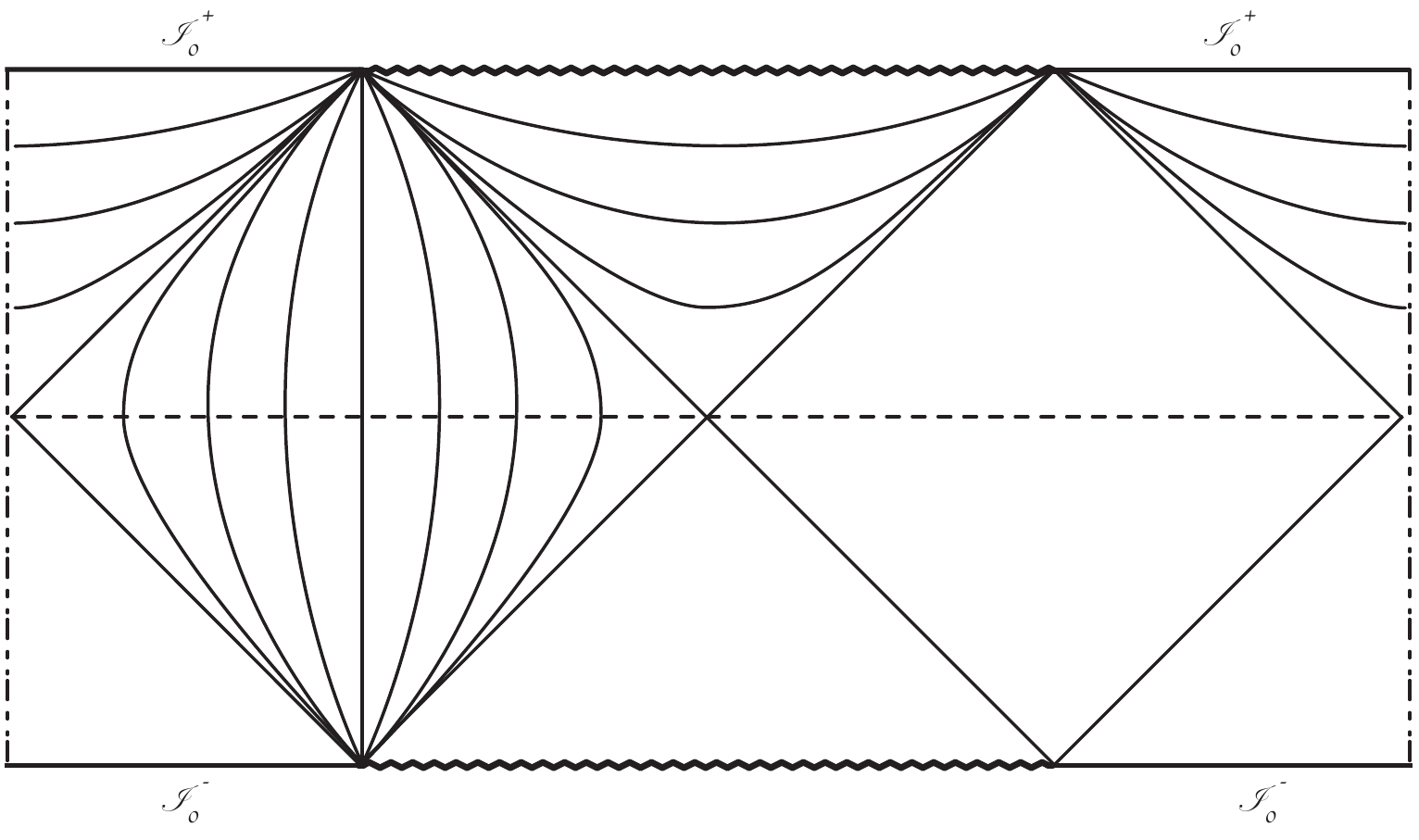}
\caption{The Penrose diagram for the $S^1\times S^2$ Schwarzschild-de Sitter spacetime with one black hole and one asymptotically deSitter region.  The dotted line is the zero extrinsic curvature Cauchy surface for this spacetime. The broken lines on the left and right side of the diagram are identified; this identification restricted to the Cauchy surface yields the $S^1\times S^2$ topology. Integral curves of the Killing vector field have been sketched in representative regions.  }
\label{fig:s1s2sds}
\end{figure}

Begin with a Cauchy surface of topology ${\mathbb R}\times S^2$. Take the metric  to be given by
\begin{equation}\label{sdsinit}{\bf  h }= \frac{1}{(\frac{2M}{r}+ \frac{\Lambda}{3}r^2)}dz^2 + r^2 d\Omega_2^2\\
\end{equation}
 where  $r(z)$ is a function of $z$ defined implicitly by
 \begin{equation}
 z(r) = \int^r_{r_h}  {dr'}\sqrt{ \frac 1{1-\frac {2M}{r'}-\frac{\Lambda}{3}r'^2}-1} \ .
 \end{equation}
 The function $z(r)$ is well defined for $0<\Lambda<\frac{1}{9M^2}$ and $r_h\leq r\leq r_c$ where $r_h$ and $r_c$ are the roots of
 ${1-\frac {2M}{r}-\frac{\Lambda}{3}r^2}$.\footnote{If $\Lambda=0$, this metric is equivalent to (\ref{sinit}).} Its inverse $r(z)$ is  well defined and positive; $r(0) = r_h$ and $r(\zeta)=r_c$  are extremal as $\frac{dr}{dz}$ vanishes at these points. Hence $r(z)$ can be smoothly extended to all values of $z$  as a periodic function in $z$. 

 The scalar curvature of (\ref{sdsinit}) is $2\Lambda$ by construction; hence this metric
 with ${\bf K}=0$ satisfies the constraints (\ref{constraints}) for the case of cosmological constant. Thus (\ref{sdsinit})  is the metric of the  time symmetric initial data set for the maximal analytic extension of Schwarzschild-de Sitter spacetime.
 
 Time symmetric initial data on the Cauchy surface with topology ${\mathbb R}\times{\mathbb R}P^2$ follows from this initial data set by antipodal identification on the $S^2$ factor of ${\mathbb R}\times{\mathbb R}P^2$ as in Example \ref{rp2}. The maximal evolution of this initial data set is clearly of the same form as that of Schwarzschild-de Sitter spacetime itself. Black hole and cosmological horizons now have topology ${\mathbb R}\times{\mathbb R}P^2$; the topology of each connected component of $\scri$ is also similarly modified.
 \begin{figure}
\includegraphics{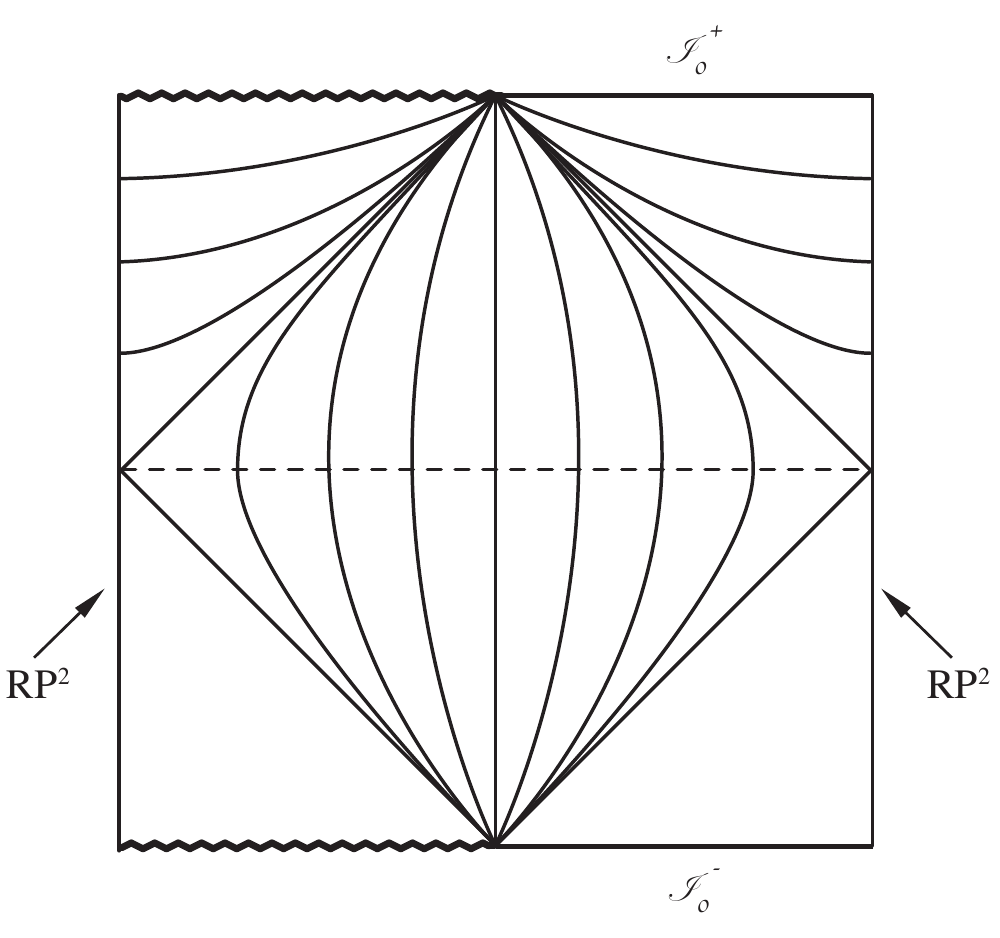}
\caption{The Penrose diagram for an ${\mathbb R}P^3\#{\mathbb R}P^3$ Schwarzschild-de Sitter spacetime with one black hole and one asymptotically deSitter region.  The dotted line is the zero extrinsic curvature Cauchy surface for this spacetime. The left line  and right line of the diagram are distinct ${\mathbb R}P^2$'s; points interior to these lines have suppressed $S^2$ factors as usual. Integral curves of the Killing vector field have been sketched in representative regions.  }
\label{fig:rp3rp3sds}
\end{figure}

The topologies $S^1\times S^2$, $S^1 \tilde \times S^2$, and $S^1\times {\mathbb R}P^2$ can be constructed from ${\mathbb R}\times S^2$ with this initial data set in a fashion similar to that of Example \ref{sc}.\footnote{The $S^1\times S^2$ identification is well known; see for example \cite{Lake:1977ui,Beig:2005ef}. However, we are not aware of the other topologies, in particular $S^1 \tilde \times S^2$, ${\mathbb R}P^3 - \{p\}$ and ${\mathbb R}P^3 \# {\mathbb R}P^3$, being discussed elsewere.} However, the identification period is no longer arbitrary, but must be an even multiple of $\zeta$, the  coordinate distance between the totally geodesic minimal and maximal 2-spheres. The evolution of an initial data set  identified with
period $2n\zeta$ will contain $n$ black hole and $n$ asymptotically de Sitter regions. 
The Penrose diagram of the $n=1$ case is shown in Figure \ref{fig:s1s2sds}. The Penrose diagrams for the $S^1 \tilde \times S^2$ and $S^1\times {\mathbb R}P^2$ cases for $n=1$ differ only in the suppressed dimensions. For the $S^1\times S^2$ and $S^1 \tilde \times S^2$ cases,  the black hole and cosmological horizons have topology ${\mathbb R}\times S^2$;  the  $S^1\times {\mathbb R}P^2$ case has horizon topology ${\mathbb R}\times{\mathbb R}P^2$.

The topologies ${\mathbb R}P^3 - \{p\}$ and ${\mathbb R}P^3 \# {\mathbb R}P^3$ are constructed similarly to that of Example \ref{rp3}. In particular ${\mathbb R}P^3 \# {\mathbb R}P^3$ can be constructed in three different ways. First one can attach two disjoint minimal surfaces, one  at $r(0)=r_h$  and the other $r(2n\zeta)=r_h$ each to a disjoint ${\mathbb R}P^2$. The resulting spacetime will have $n$ asymptotically de Sitter regions, $n-1$ black holes and two black holes of  ${\mathbb R}P^3$ geon type. Second, one can carry out the same construction with  two disjoint maximal surfaces, one at $r(0)=r_c$  and the other at $r(2n\zeta)=r_c$; the evolution of such initial data will be a spacetime with
$n$ black holes, $n-1$ asymptotically de Sitter regions and two asymptotically de Sitter regions, each with a connected component of $\scri$ of topology ${\mathbb R}P^3-\{p\}$.
 Third, one can do so with one minimal surface at $r(0)=r_h$, and one maximal surface at $r((2n+1)\zeta)=r_c$.  The resulting spacetime has
 $n-1$ black holes, $n-1$ asymptotically de Sitter regions, one black hole of  ${\mathbb R}P^3$ geon type and one asymptotically de Sitter region with a connected component of $\scri$ of topology ${\mathbb R}P^3-\{p\}$.  The Penrose diagram for the $n=1$ case is given in Figure \ref{fig:rp3rp3sds}. As for the ${\mathbb R}P^3$ geon, there is a trapped set of topology ${\mathbb R}P^2$ on the intersection of the Cauchy surface with the ${\mathbb R}P^3$ geon type black hole horizons that is not a trapped surface.

\end{exam}

Initial data sets  for Schwarzschild-de Sitter cosmologies have the same form as those in Example \ref{sc} but again the situation is more complex than in the Schwarzschild case. 
\begin{exam}(Schwarzschild-de Sitter Cosmologies) \label{sdsc}

Choose initial metric  (\ref{hdata}) and extrinsic curvature (\ref{kdata}) on a Cauchy surface of closed global topology such as
$S^1\times S^2$,
 $S^1{\tilde { \times }} S^2$, $S^1\times {\mathbb R}P^2$ and ${\mathbb R}P^3\#{\mathbb R}P^3$.
This initial data trivially satisfies the momentum constraint, but the hamiltonian constraint is now
\begin{equation}\label{hc2}
\frac 1{a^2} + \frac {2bc}{a^4} + \frac {c^2}{a^4} = \Lambda \ .
\end{equation}
Thus, as for Example \ref{sc}, one of the three parameters, $a,b,c$ is determined in terms of the other two.
This evolution of this initial data set yields the interior form of the Schwarzschild-de Sitter solution, 
\begin{equation}
ds^2 = -\frac{dt^2}{\left( \frac \Lambda{3} t^2 + \frac {2M}{t} - 1\right) } + \left(\frac \Lambda{3} t^2 + \frac {2M}{t} - 1\right)dr^2 + t^2d\Omega _2^2 \ , \label{ssmetricinside2}
\end{equation}
 where now
\begin{equation} 
M= \frac {\Lambda}{3} a^2 - \frac{bc}a
\end{equation}
 $ dr= \alpha d\psi$ and with scaling $\alpha$ chosen
so that $(\frac{\Lambda}{3} a^2 +\frac{2M}{a} - 1)\alpha^2 = a^2$.  

As in Example \ref{sc}, the above  solutions explicitly  exhibit a spacelike translational Killing vector.
However, unlike the Schwarzschild case, whether or not the spacetime is singular to both the future and past depends on the parameters. If $bc<0$, then the evolution of this initial data yields the black or white hole interior solution of Schwarzschild-de Sitter spacetime. Its behavior is qualitatively the same as that of Example \ref{sc};  Cauchy surfaces with closed spatial topology are singular to both the past and future. If $bc>0$ and $b>0$, the evolution instead yields a solution that is future asymptotically de Sitter  and singular to its past. Conversely, if  $b<0$  then it is past asymptotically de Sitter  and singular to its future. These two cases never exhibit a static region in their evolution.

If $a,b,c$ are such that $M=0$, then the solution is an identification on the Kasner slicing of de Sitter spacetime
(see  \cite{MorrowJones:1993zu, Schleich:2008zr}) \begin{equation}\label{kasnerslice} ds^2 = -dt^2 + H^2 \left( \sinh ^2 \left( t/H \right) d\psi ^2
	+ \cosh ^2 \left( t/H \right) d\Omega _2 ^2
			\right)\  \end{equation}	
	where  $H= \sqrt{\frac{3}{\Lambda}}$.
As the spacetime is de Sitter, it is locally static even though this form of the metric does not explicitly exhibit such a locally static Killing vector field. These closed locally de Sitter cosmologies are either singular to the past and asymptotically de Sitter to the future of the Cauchy surface or the reverse as determined by the extrinsic curvature.		

Finally, if $c=0$,  then this initial data is that for the Nariai solution in the form \begin{equation}
ds^2 = -\frac{dt^2}{( \Lambda t^2-1)}  + {( \Lambda t^2-1)} {dr^2} + 
\frac 1{\Lambda} d\Omega _2^2\  .  \label{nariaimetric2}
\end{equation}
Notably, the maximal extension of the evolution of Nariai initial data for closed topology are also no longer singular either to the past or future. Again, the spacetime is locally static, though  (\ref{nariaimetric2}) does not explicitly exhibit this property. 
\end{exam}
Clearly, although initial data for closed cosmologies with positive cosmological constant can result in spacetimes with qualitatively the same behavior as those for the vacuum case, it also results in spacetimes with qualitatively different behavior. The extrinsic curvature determines the behavior of the solution.

Examples \ref{rp2}-\ref{sdsc} have particularly simple initial data; in particular, their extrinsic curvature takes a particularly simple form. However it is clear that similar techniques can be used to construct more general initial data sets  analogous to the  constant mean curvature slices for Schwarzschild \cite{brill,Beig:1997fp,Malec:2003dq} and Schwarzschild-de Sitter spacetimes \cite{Nakao:1990gw,Beig:2005ef}.

\section{Locally Spherically Symmetric Spacetimes}

We next turn to initial data sets for spacetimes that are locally but not globally spherically symmetric. A particularly important set of such spacetimes are those that are locally isotropic ( \cite{Wolf} ch. 12):

\begin{defn} \label{lss4} A manifold, $M$, with lorentzian metric, $g_{ab}$,
is locally isotropic if there is a local isometry that maps any two tangent vectors $X$, $Y$ of equal norm into each other about every point. \end{defn}
Well known examples of such spacetimes are those built from identifications on globally isotropic Minkowski and de Sitter spacetimes.  
Indeed, the folklore in general relativity is that all locally isotropic spacetimes are identifications on globally isotropic ones. However, this belief is false for the locally de Sitter case;  one can construct initial data sets that evolve to locally  de Sitter spacetimes whose covering space is not  de Sitter spacetime itself. This was first shown by  Morrow-Jones and Witt \cite{MorrowJones:1988yw,MorrowJones:1993zu}. They constructed locally spherically symmetric initial data sets for de Sitter spacetimes with  generic  topology. 
Birkhoff's theorem then implies that that the evolutions of these initial data sets are locally de Sitter spacetimes; hence they are locally isotropic. They then give  examples of initial data sets that are locally de Sitter spacetime but whose maximal evolution is not de Sitter spacetime itself.  A spacetime based construction of  these solutions was  presented in 
\cite{Bengtsson:1999ia}; the results of  \cite{MorrowJones:1993zu} in 3+1 dimensions were then later reproduced without reference in \cite{student}.   A simple characterization of the behavior of such locally spherically symmetric initial data sets was given in \cite{Schleich:2008zr} by proving that the global dynamics of certain locally isotropic de Sitter spacetimes, designer de Sitter spacetimes, are in fact not determined by a single scale factor.  

In this Section we begin by constructing the simple examples of locally isotropic spacetimes that are simply identifications on the globally isotropic ones. We then construct a family of examples of de Sitter initial data sets, whose maximal evolution is not de Sitter spacetime itself.

\begin{exam}(Identified Minkowski  spacetimes)\label{idmink}
Take the initial data on the Cauchy surface ${\mathbb R}^3$  to be a flat metric ${\bf h} = dx^2+dy^2+dz^2$ with ${\bf K} =0$. The identification ${\mathbb R}^3 /({\mathbb Z}\times{\mathbb Z}\times{\mathbb Z})$ where the map is given by
$(x,y,z)\to (x+m\alpha, y+n\beta, z+p\gamma)$, $(m,n,p)$ integers yields a locally spherically symmetric initial data set on the 3-torus. This initial data set and the evolved spacetime are not globally spherically symmetric as the local Killing vectors about any point cannot be extended to form a smooth global Killing vector field.  Similar constructions with the identification generated by maps that are a composition of  translations and rotations yield flat, locally spherically symmetric initial data sets on 9 other distinct closed 3-manifolds \cite{Wolf}.\footnote{This construction is also reviewed in \cite{Levin:2001fg}.}  The evolutions of these initial data sets are locally Minkowski spacetime, and therefore are locally isotropic. Their global topologies are $ F \times {\mathbb R}$ where $F$ is one of the 10 distinct closed 3-manifolds admitting a flat metric.

A similar construction produces locally spherically symmetric initial data sets on closed hyperbolic 3-manifolds. Take spherically symmetric initial data ${\bf h} = dr^2+\sinh^2 r d\Omega_2^2$ 
and ${\bf K} ={\bf h} $ on ${\mathbb R}^3$. It is easy to verify that the constraints (\ref{constraints}) are satisfied by this initial data. Clearly it is a partial Cauchy surface for Minkowski spacetime. This surface lies in the interior of the future light cone of a point; the light cone itself is the Cauchy horizon for the evolution of this initial data set. This initial data set is manifestly globally spherically symmetric and isotropic. Now, 
initial data sets on closed hyperbolic 3-manifolds with topology $\Sigma = {\mathbb R}^3/\Gamma$ can be formed from this one by taking its quotient with a discrete subgroup $\Gamma$ of the isometry group of the hyperbolic space that acts freely and properly discontinuously. After such an identification, of course, the initial data set  and its evolution are now only locally spherically symmetric.  In addition, the locally isotropic  spacetime resulting from the evolution of this initial data set, though locally isometric to Minkowski spacetime,  can no longer be extended across the Cauchy horizon; the spacetime is singular to the past. 
\end{exam}
De Sitter initial data sets that are locally, but not globally spherically symmetric can be constructed by the same method as in Example \ref{idmink}. 
\begin{exam}(Identified de Sitter spacetimes)
The three Robertson-Walker forms of de Sitter spacetime are
\begin{equation}\label{sphere} ds^2 =  -dt^2 + H^2 \cosh (t/H )
	\left( d\psi^2 + \sin^2 \psi
	d\Omega _2 ^2  \right)\end{equation}
	with Cauchy slice $ S^3$,
\begin{equation}\label{flat} ds^2 = -dt^2 +H ^2 e^{ 2 t/H }
	\left( d\psi ^2 + \psi ^2
	 d\Omega _2 ^2  \right)\end{equation}
	  with partial Cauchy slice $ {\mathbb R^3}$ and
\begin{equation} \label{hyp} ds^2 = -dt^2 + H^2 \sinh ^2 \left( t/H \right)
	\left( d\psi ^2 + \sinh ^2 \psi 
	 d\Omega _2 ^2 \right)\end{equation}
	also with partial Cauchy slice ${\mathbb R^3}$. Constant $t$ slices of these spacetimes yield  initial data sets that can  be written in the form
\begin{align} {\bf h} &=
	a^2\left( d\psi^2 +\sin^2\psi
	d\Omega _2 ^2  \right) \nonumber\\
	{\bf K} &= \pm\sqrt{\frac{\Lambda}{3} - \frac 1{a^2}} {\bf h}\ ,\label{inits} \\
	 {\bf h} &=
	a^2\left(d\psi^2 +\psi^2
	d\Omega _2 ^2\right) \nonumber\\
	{\bf K} &=  \pm\sqrt{\frac{\Lambda}{3} }{\bf h}\ ,\label{initflat}\\
	{\bf h} &=a^2 \left( d\psi^2 +\sinh^2\psi
	d\Omega _2 ^2  \right)\nonumber\\
	{\bf K} &=  \pm \sqrt{\frac{\Lambda}{3} + \frac 1{a^2}}  {\bf h}\label{inithyp}
	\end{align}
	respectively.  Identifications of the form $S^2/\Gamma$ on the initial data set (\ref{inits}), where $\Gamma$ is a freely acting finite subgroup of $O(4)$, the isometry group of $S^3$, yield locally spherically symmetric initial data sets.  When $\Gamma={\mathbb Z}_2$, this space is ${\mathbb R}P^3$ and is globally spherically symmetric. Those of the form ${\mathbb R}^3/\Gamma$ on initial data of form (\ref{initflat}) for suitable discrete group $\Gamma$ will yield the ten closed flat 3-manifolds discussed in Example \ref{idmink}. Those of the form ${\mathbb R}^3/\Gamma$ on initial data sets of form (\ref{inithyp}) yield smooth  initial data sets on hyperbolic 3-manifolds.
Clearly, all these these initial data sets evolve to form locally de Sitter spacetimes. 
Note that for identifications on the flat and hyperbolic initial data sets, these spacetimes in general will be singular and inextendible either to the past or future due to their nontrivial topology.  De Sitter spacetimes arising from this type of initial data result in observable consequences in the cosmic microwave background \cite{Stevens:1993zz,Cornish:1997rp,Cornish:1997hz,Weeks:1998qr,Levin:2001fg,Reboucas:2004dv,Luminet:2005tn}. Predictions based on such models have yielded constraints on the  topology of the universe. 

\end{exam}

The next example is  a class of designer de Sitter spacetimes:
\begin{defn} \label{lss5} A designer de Sitter spacetime is any locally de Sitter spacetime with Cauchy slice of
the form $\Sigma = \Sigma _1\# \Sigma _2\# \Sigma _3\#   \dots   \#\Sigma _k\# \dots \ \  $
where  at least two $\Sigma _i$'s  are not $S^3$'s or at least one $\Sigma_i$ is ${\mathbb R}\times S^2$ or an identification on ${\mathbb R}\times S^2$.  \end{defn}
Observe that $S^3$ is the identity under the connected sum: $\Sigma\# S^3= \Sigma$ for any $\Sigma$. Therefore definition \ref{lss5} requires two of the factors to be other than an $S^3$ to ensure $\Sigma$ not have a simple topology. 
The construction below uses the techniques introduced in  \cite{MorrowJones:1993zu}.

\begin{exam}(Simply connected designer de Sitter spacetimes)\label{designer}
These examples  are all formed from one locally spherically symmetric initial data set on ${\mathbb R}\times S^2$: Let
\begin{align}{\bf  h}&=d\psi^2 + \gamma^2d\Omega ^2\nonumber \\
{\bf K}&=\alpha d\psi^2 +\gamma^2\beta d\Omega ^2\label{ssinitial2}\end{align}
where $\gamma$,  $\alpha$, and $\beta$ only depend on $\psi$. The constraints (\ref{constraints}) are then
\begin{align}
\frac {1}{2} R + 4\alpha\beta +\ \beta^2 &= \Lambda\nonumber\\ 
\partial_\psi \beta + \frac{\partial_\psi \gamma}{\gamma} \beta &= \frac{\partial_\psi \gamma}{\gamma} \alpha 
\end{align}
where 
\begin{equation}R = \frac {-4\gamma\partial^2_\psi \gamma - 2 (\partial_\psi\gamma)^2 +2}{\gamma^2}\ .\nonumber\end{equation}
The locally isotropic solution of these equations is 
\begin{align}\beta^2& = \frac {\Lambda}{3} + \frac{(\partial_\psi\gamma)^2 - 1}{\gamma^2}\\
\alpha &= \frac1{\beta}\left( \frac {\Lambda}{3}+ \frac {\partial^2_\psi \gamma}{\gamma}\right)\label{Ksln}
\end{align}
and is well defined for any function $\gamma$ such that (\ref{Ksln}) are real, bounded functions everywhere. 

Define the bump function
\begin{equation}
f(x) = \frac {\int_{-\infty}^x dx' \phi(x')\phi(\psi_0- x')}{\int_{-\infty}^\infty dx' \phi(x') \phi(\psi_0 - x')}
\end{equation}
 where $\phi$ is the smooth function
\begin{equation}
\phi(x)=
\begin{cases}
0&x\leq 0\\
\exp(-\frac 1{x^2})& x>0\ .\\
\end{cases}
\end{equation} This bump function is smooth,  vanishes for $x<0$ and monotonically increases to $1$ for $x>\psi_0$. 
Take $\gamma$ to be 
\begin{equation}\label{gamma}
\gamma=
\begin{cases}
\psi&\psi\geq \psi_0\\
\psi f(\psi) + A(1-f(\psi))&0<\psi<\psi_0\\
A&\psi \leq 0\\
\end{cases}
\end{equation}
where $A$ is a constant $A>H$ and $\psi_0> A$. This choice yields
\begin{equation}
\begin{cases}
\beta=\frac 1{H}\ \ \ \ \ \ \ \ \ \ \ \ \ \ \ \ \ \ \  \alpha=\frac 1{H}&\psi\geq \psi_0\\
\beta=\sqrt{\frac{1}{H^2}-\frac 1{A^2}}\ \ \ \ \ \ \ \ \alpha=\frac 1{H \sqrt{1-\frac {H^2}{A^2}}}&\psi \leq 0\\
\end{cases}
\end{equation}
with both $\alpha$ and $\beta$ smooth, real functions interpolating between these values in $0<\psi<\psi_0$. It is apparent that for $\psi<0$, the initial data is that for the Kasner form of de Sitter (\ref{kasnerslice}) and for $\psi>\psi_0$, that for the flat slicing (\ref{initflat}). Thus
(\ref{Ksln}) with $\gamma$ given by (\ref{gamma})  geometrically interpolates between these two solutions. As both ends of this initial data set are locally de Sitter spacetimes, the analyticity of spacetimes satisfying Birkhoff's theorem guarantees that this initial data set evolves to a locally de Sitter spacetime everywhere. This metric is illustrated in Figure \ref{fig:neck}.  
\begin{figure}
\includegraphics{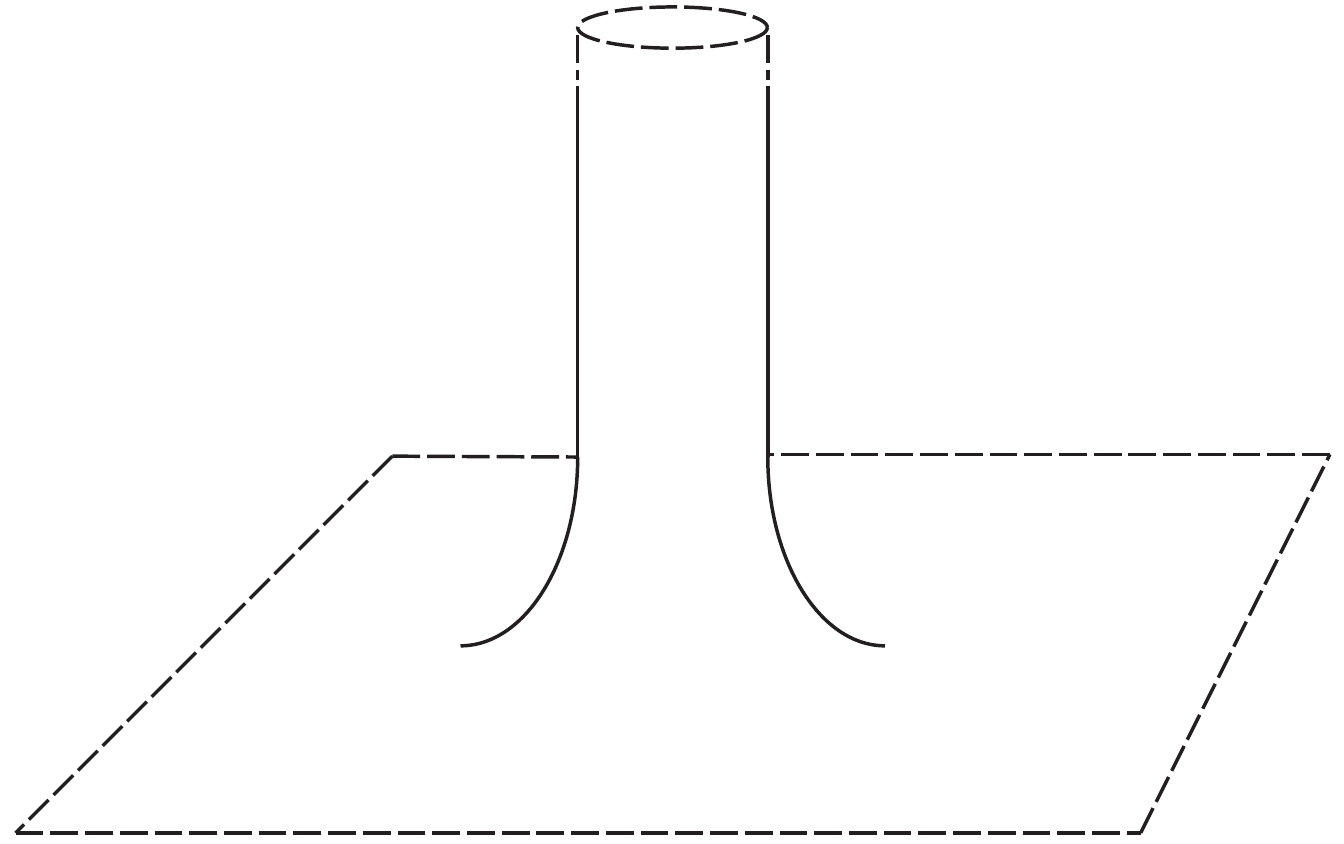}
\caption{ The geometry of a ${\mathbb R}\times S^2$ metric that interpolates from a flat metric to the cylindrical one. One dimension is supressed. Identification on the boundaries yields the $T^3- B^3$ example. This illustration clearly describes the geometry in the  fundamental cell in the covering space of the  $T^3- B^3$ initial data set. }
\label{fig:neck}
\end{figure}

Iteration of this construction yields a countably infinite set of locally de Sitter spacetimes.  For example, pick $n$ constants $\psi_n$ and $A_n$ such that $\psi_n>A_n>H$ and pick $n$ points on ${\mathbb R}^3$ with initial data (\ref{initflat}), each separated from all others by a distance greater than $2\bar \psi$ where $\bar \psi$ is the supremum of $\psi_n$. In a suitably sized neighborhood of each point, replace the existing data with that of (\ref{Ksln}) with the obvious choice of constants. The resulting initial data has topology ${\mathbb R}^3 - nB^3$, that is  ${\mathbb R}^3$ minus $n$  3-balls.  A case with $n=4$ is given in Figure \ref{fig:necks}.
\begin{figure}
\includegraphics{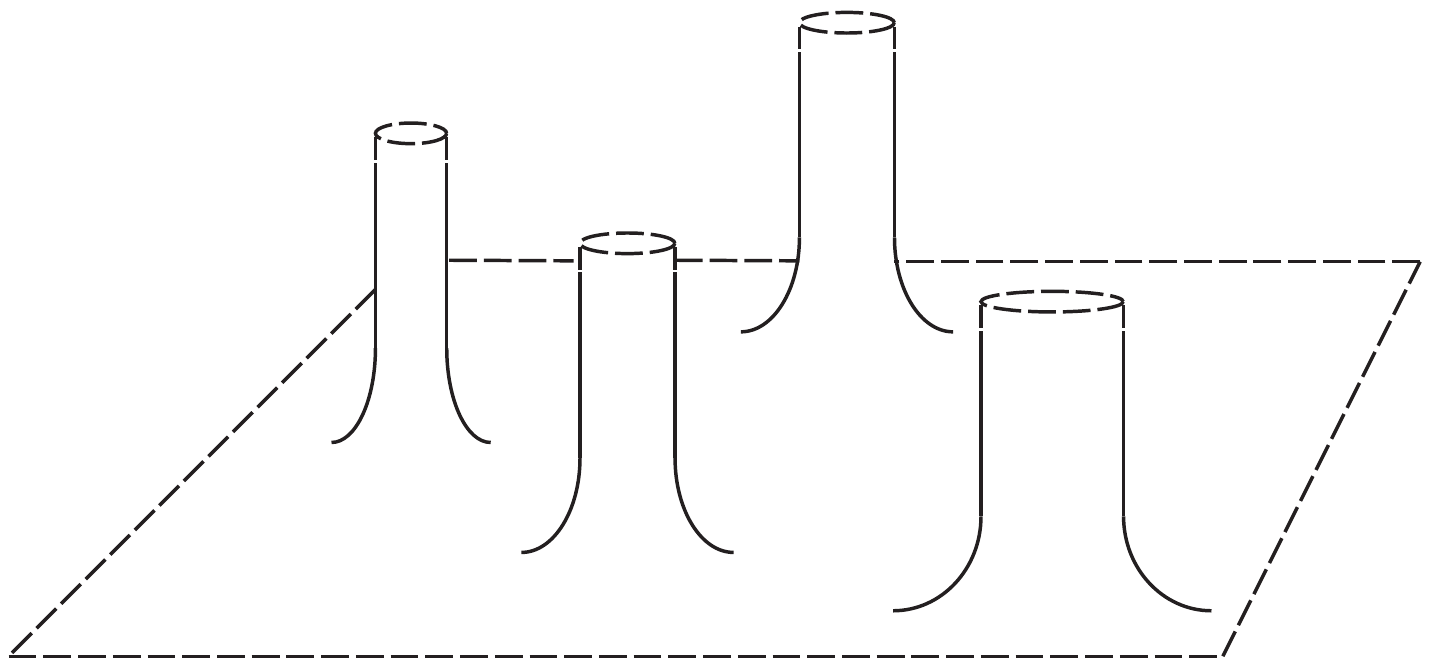}
\caption{ The geometry of four ${\mathbb R}\times S^2$ necks attached to ${\mathbb R}^3$ through the geometric construction of Example \ref{designer}. One dimension is supressed. Note that the spacing and diameter of the necks are not uniform; these are parameters as described in text.  }
\label{fig:necks}
\end{figure}

Another simple example is given by taking initial data (\ref{Ksln}) and, noting that by a straightforward coordinate change, ${\bf h} = dx^2 + dy^2 + dz^2$ outside of $\psi=\psi_0$. Periodic identification of planes at $x=\pm L$, $y=\pm L$, $z=\pm L$ results in an initial data set on the topology $T^3-B^3$. The universal covering space for this manifold \footnote{To construct the universal covering space ${\cal M}$ of $M$ pick a point $x_0\in
M$ and consider the set of smooth paths $P=\{c:[0,1]\rightarrow M|c(0)=x_0\}$.  A
projection map $\pi:P\rightarrow M$ is defined by  $\pi(c(t))=c(1)$.  Let ${\cal
M}$ be $P$ modulo the equivalence relation, $c_1\sim c_2$ if and only if
$c_1(1)=c_2(1)$ and $c_1$ is homotopic to $c_2$ with endpoints fixed. The
projection map $\pi$ is then well defined and smooth as a map $\pi:  {\cal
M}\rightarrow M$. }
 has countably infinite second homology. This is easy to see; the covering space has a generator of $H_2({\mathbb Z})$ for each $L\times L\times L$ cell in the universal cover,  ${\mathbb R}^3$,  of the 3-torus.
 
One can also attach a second ${\mathbb R}^3$  by moving a distance $D$ away from $\psi=0$ along the ${\mathbb R}\times S^2$ gemetrical factor then attaching on this neck to a second plane with flat initial data  utilizing  (\ref{Ksln}). This initial data set still has topology ${\mathbb R}\times S^2$; however one can now attach an arbitrary number of planes together with an arbitrary number of necks by iteration of this construction and that of adding a neck to a plane described above. Figure \ref{fig:planes} is an example of a geometry. If two or more necks connect the same two planes, the resulting space will not be simply connected. However, its universal cover will be  an example of a simply connected designer de Sitter initial data set. 
\begin{figure}
\includegraphics{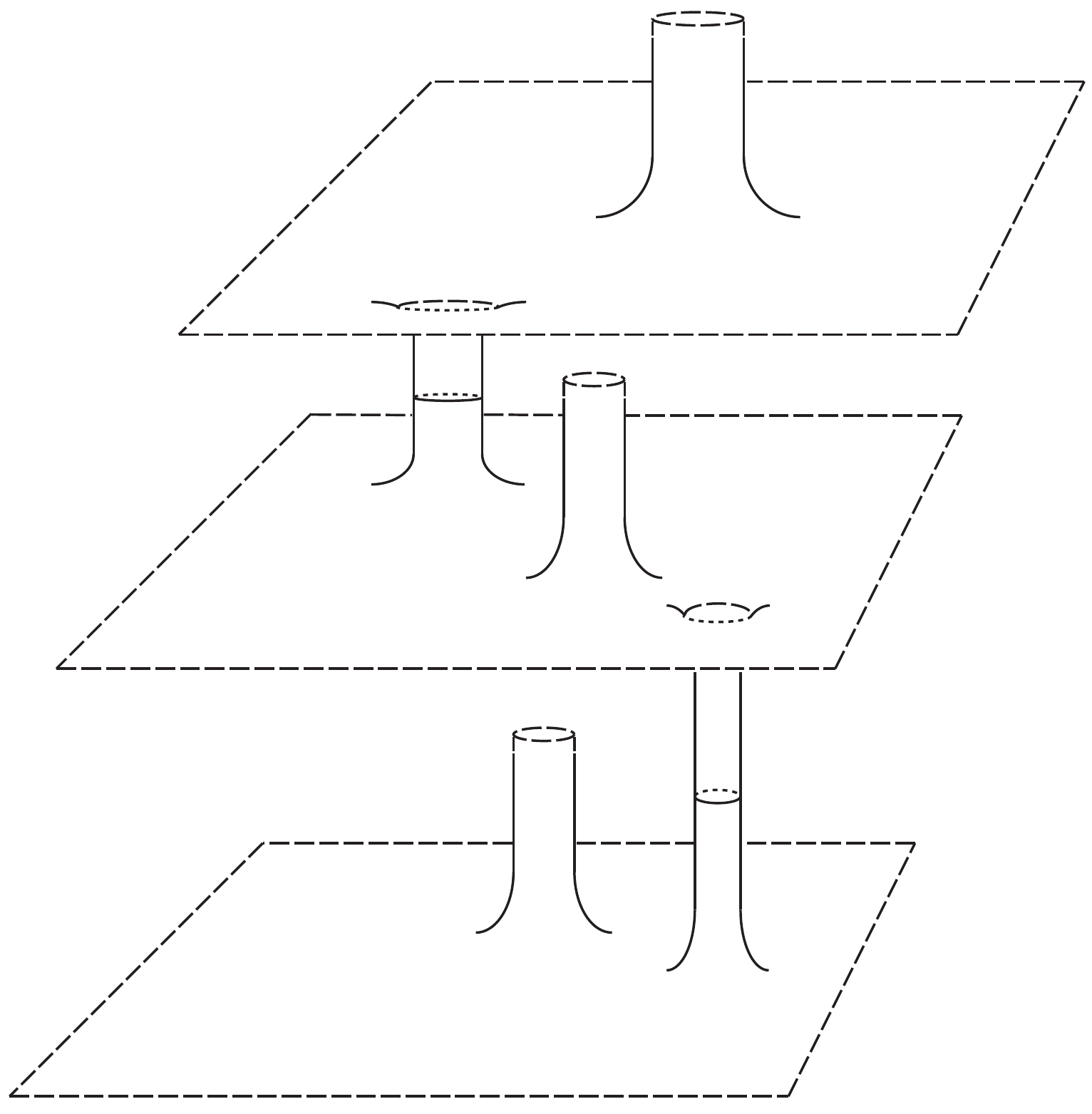}
\caption{ The geometry of three  ${\mathbb R}^3$ planes with flat metric attached together by ${\mathbb R}\times S^2$ necks  with three more ${\mathbb R}\times S^2$ necks additionally attached. Note that this figure is not a faithful representation of this geometry as  the three additional necks extend to infinite length; this feature cannot be realized in a diagram embedded in Euclidean space. Again observe that the spacing and diameter of the necks  are specifiable parameters. }
\label{fig:planes}
\end{figure}

More complicated examples can be constructed by similarly attaching  flat and hyperbolic ${\mathbb R^3}/\Gamma$ initial data sets, spherical $S^3/\Gamma$ initial data sets and $({\mathbb R}\times S^2)/\Gamma$ initial data sets in a local, spherically symmetric neighborhood using bump functions. Their universal covers will be simply connected designer de Sitter initial data sets. There is no restriction on the number of factors in these connected sums.
Thus there is a countably infinite number of  parameters in these sets of designer de Sitter spacetimes:
those set by the global topology $\Sigma$ and others characterizing geometrical factors such as relative positions of the gluing regions in the geometrization of the connected sum.  Such simply connected initial data sets with a sufficient number of factors cannot be isometrically embedded in de Sitter spacetime \cite{MorrowJones:1993zu,Bengtsson:1999ia}.

\end{exam}

\section{Application of Birkhoff's theorem in the analysis of initial data sets}\label{section6}
It is important to note that Theorem \ref{init} and Birkhoff's theorem also provide a simple method of analysing spherically symmetric initial data sets whose form is not immediately recognizable. This has already been used in the identification of locally de Sitter spacetimes as the evolution of locally spherically symmetric initial data sets in \cite{MorrowJones:1993zu}. We give two additional examples below. First we present an abstract, globally spherically symmetric initial data set on  ${\mathbb R}P^3-\{p\}$ and then invoke Birkhoff's theorem to recognise the associated spacetime as the ${\mathbb R}P^3$ geon. 

\begin{exam}( The ${\mathbb R}P^3$ geon  II) \label{rp3abs}

First abstractly construct a geodesically complete, spherically symmetric metric on  ${\mathbb R}P^3 - \{p \}$ using a procedure suggested by Geroch \cite{rg}. Let $h_{ab}$ be
 the round metric on 
${\mathbb R}P^3 $ and ${\delta }_p$ be the delta function at the point $p$. Define the operator $L=-8{ D}^2 + { R}$ where the covariant derivative and the scalar curvature are  with respect to  $h_{ab}$. As the operator
$L$ is positive,  it is invertible. Hence  
\begin{equation} L\phi = (-8{ D}^2 + { R}) \phi ={\delta }_p \label{cil}
\end{equation}
  has a unique positive 
solution; it is the Green's function for $L$ on ${\mathbb R}P^3$.\footnote{Note that some definitions of the Laplacian on a closed manifold subtract the volume from the operator. We do not do so here.}  Moreover, the equation $L\phi ={\delta }_p$ is invariant under $SO(3)$. Therefore $\phi $ is also invariant; it is a spherically symmetric function.

Now, define the new metric
${\bar h}={\phi }^4h$. By (\ref{cil}),
the scalar curvature of the metric ${\bar h}_{ab}={\phi }^4h_{ab}$ vanishes,
\begin{equation}{\bar R}={\phi }^{-5}\bigl(-8{ D}^2\phi + {R}\phi \bigr)=0\ ,\end{equation} 
at all points except $p$. At $p$, the Green's function $\phi $ has a pole.  This pole "opens up"  the geometry ${\bar h}_{ab}$ on ${\mathbb R}P^3 - \{p \}$, that is ${\bar h}_{ab}$ is an asymptotically flat,  geodesically complete, spherically symmetric metric.

To see how this works, consider the limiting case where $R\to 0$, that is the  case in which the radius of the ${\mathbb R}P^3$ becomes very large. Pick a spherically symmetric coordinate system in a neighborhood centered at $p$; 
$h_{ab}$  can then be written to lowest order to higher order in Riemann normal coordinates as
\begin{equation*} {\bf h} = dr^2 + r^2 d\Omega_2^2 \ . \end{equation*}
As $\phi$ is spherically symmetric, (\ref{cil})  becomes
\begin{equation}\label{agf}
\frac 1{r^2}\partial_r(r^2\partial_r \phi (r)) = 0
\end{equation}
with presence of the delta function source equivalent to the condition
\begin{equation*}
\int_S d\sigma \partial_r(r^2\partial_r \phi (r)) =  1 \ 
\end{equation*}
where $S$ is a sphere of constant radius and $d\sigma$ the measure on the unit sphere.
The solution to (\ref{agf}) is simply  the Green's function  $\phi(r) = \frac 1{4\pi r}$. Thus  $\bar h_{ab}$ explicitly takes the form
\begin{equation} {\bf h} = \left(\frac 1{4\pi r}\right)^4\left(dr^2 + r^2 d\Omega_2^2\right) \end{equation}
in this neighborhood of $p$. The coordinate transformation $r' = 1/({16\pi^2r})$ shows that
\begin{equation} {\bf \bar h} = \left(dr'^2 + r'^2 d\Omega_2^2\right) \end{equation}
as $r' \to \infty$. Hence $\bar h_{ab}$ is flat in a neighborhood of $p$ to leading order.

This limiting behavior characterizes the leading order behavior of $\bar h_{ab}$ for any metric with nonnegative scalar curvature on $RP^3$. This can 
be seen explicitly by repeating the above procedure in Riemann normal coordinates for the general case. The leading term in $\phi$ will be the same;  the higher order terms in the expansion result in a mass term. Consequently  $\bar h_{ab}= (1+M/(2r))^4\delta_{ab} $ in isotropic coordinates in a neighborhood of infinity.  The resulting metric $\bar h_{ab}$ will be asymptotically flat and spherically symmetric; the geometry of ${\mathbb R}P^3$  produces the mass of the asymptotically flat spacetime.\footnote{Note that if this calculation is done on the covering space, it is necessary to solve (\ref{cil}) with not one but two delta functions, corresponding  to punctures at the north and south poles of $S^3$. The calculation with one delta function on $S^3$ yields the flat metric.}

 Given that ${\bar h}_{ab}$ is an asymptotically flat,  geodesically complete, globally spherically symmetric metric  with $\bar R= 0$, the choice $K_{ab} = 0$ clearly satisfies the constraints. Hence this initial data on ${\mathbb R}P^3 - \{p\}$ is globally spherically symmetric and its evolution is a spacetime that is locally Schwarzschild.
 
It is easy to find this initial data explicitly in a familiar form; the cover of this Cauchy slice is
${\mathbb R}\times S^2$, the topology of the Cauchy surface of the maximally extended Schwarzschild spacetime. Hence Birkhoff's theorem implies that the cover of the initial data set is simply that for the time symmetric slice of maximal analytic extension of Schwarzschild itself. 
\end{exam}

Other examples can be constructed by similar procedures when the  isometry groups of the topology can be found rigorously, providing a basis for the result without explicit computation.
The next example violates the strict fall-off conditions of asymptotically flat data; however, it is still
globally spherically symmetric. Birkhoff's theorem is utilized to identify the spacetime resulting from its evolution.

\begin{exam}(An initial data set on  ${\mathbb R}^3$ with negative ADM mass) \label{fall-off}

General globally spherically symmetric initial data sets can be constructed using a 
simplified version of the constraints  from \cite{Witt:1986ng,Witt:2009za}. 
First observe a general spherically symmetric metric and extrinsic curvature can be written  on the Cauchy surface ${\mathbb R}^3$ 
as 
\begin{align}{\bf  h}&=\xi dr^2 +r^2d\Omega ^2\nonumber \\
{\bf K}&=\alpha dr^2 +\beta r^2d\Omega ^2\label{ssinitial}\end{align}
where $\xi$, $\alpha$, and $\beta$  only depend on $r$. The constraints (\ref{constraints}) for the vacuum case are then
\begin{align}
0& =\frac{1} {16\pi}\Bigl[{\frac{2\xi ^{-2}\xi'} r} +{\frac{2(1-\xi^{-1})} {r^2}}+4\alpha \beta \xi ^{-1}+2\beta ^2 \Bigr]\label{rhoeqn}\\
0&={\frac1{4\pi}}\Bigl[-\beta '\xi ^{-1}+{\frac{\xi ^{-1}} r}(\alpha \xi ^{-1}-\beta )\Bigr]\ .
\end{align}
This implies
\begin{equation} \alpha =\xi (\beta 'r+\beta )\ .\end{equation} 
Substitution into (\ref{rhoeqn})  results in
\begin{equation}
\label{ssrho}
\rho = \frac{1}  {8\pi r^2}  \frac{d} {dr}\Bigl[\beta ^2r^3-\xi^{-1}r+r\Bigr]\ .\end{equation}
 Taking the metric to be flat at the origin, $r=0$, and integrating (\ref{ssrho})
from $0$ to $r$ yields 
\begin{equation}
\label{integrated}
\beta ^2r^3-(\xi^{-1} -1)r=0 
\end{equation}
for everywhere vacuum initial data.
Pick $\xi^{-1}= 1-\frac {2M(r)}{r}$
where $M(r)$ is given by
\begin{equation}
\label{Afunction}
M(r)=
\begin{cases}
M_0 & r\geq r_3\\
\text{smoothly monotonic }&r_2<r<r_3\\
0& r\leq r_2\\
\end{cases}
\end{equation}
where $M_0$ is a negative constant constant, $M_0<0$ and $r_2$ and $r_3$ are positive constants with $r_2<r_3$. 
It immediately follows from (\ref{integrated}) that
\begin{equation}\beta ^2=\frac{(\xi^{-1} -1)}{ r^2}=  \frac{-2M(r)}{ r^3}\ .
\end{equation}
Note that $\beta $ is well defined as $M(r)\leq 0$ for all $r$. Similarly, 
 $\alpha $ is also well defined as $\beta$ is a smooth function and $\xi^{-1}$ is a smooth function that is nonvanishing everywhere.
 
 The metric $h_{ab}$ is geodesically complete and $K_{ab}$ is smooth and well defined everywhere. Hence this initial data set is well posed. For $r<r_2$, ${\bf h} = dr^2 + r^2 d\Omega_2^2$ and ${\bf K} = 0$; it is initial data for a piece of Minkowski spacetime. For $r>r_3$,
${\bf h} = \frac{dr^2}{1-2M_0/r} + r^2 d\Omega_2^2$, the same spatial metric as found on the time symmetric slice of
negative mass Schwarzschild. However, this initial data set is not that for negative mass Schwarzshild as the extrinisic curvature is nonvanishing; ${\bf K} = \sqrt{\frac {-2M_0}{r^3}} ( -\frac{r}{2r-4M_0} dr^2 + r^2d\Omega_2^2)$ which has fall-off of $1/r^{\frac 32}$ as $r\to\infty$. In the region $r_2<r<r_3$ the initial data smoothly interpolates between these two sets.

A simple computation shows that the ADM mass\footnote{The ADM mass is \begin{equation}
M_{ADM} = \lim_{r\to\infty} \frac{1}{16\pi}\int_{S_r^2}d\sigma^a  \sum_b
( \partial_bh_{ab} - \partial_a h_{bb} ) 
\end{equation}
which is the limit of the surface integral over  a large 2-sphere in the Cauchy surface as its radius goes to infinity.}
 of this initial data set is negative, $M_{ADM}=M_0<0$.
Hence we seemingly have constructed regular initial data for a negative mass spacetime, in contradiction to the positive mass theorem  \cite{Schon:1979rg, Schon:1981vd, Witten:1981mf}.

Birkhoff's theorem allows one to readily identify the spacetime corresponding to this initial data set.  By Birkhoff's theorem, this spacetime must be locally Schwarzschild; furthermore, it must be analytic.  Observe that its evolution in a neighborhood of the origin is Minkowski spacetime, i.e. Schwarzschild spacetime with zero mass parameter $M$. As the spacetime is analytic, this mass parameter cannot change. Consequently, this evolution of the entire initial data set must have zero mass parameter; therefore it evolves to a locally Minkowski spacetime. Equivalently, the Weyl curvature  of the evolution of the initial data vanishes in an open neighborhood of the origin and, as the spacetime is analytic,  it therefore vanishes everywhere. Again we conclude that spacetime is locally Minkowski spacetime. 

So how did a piece of Minkowski spacetime acquire a negative mass? The answer is that it is not an asymptotically flat initial data set. $K_{ab}$ does not obey the required  fall-off conditions;
It falls-off  as $1/{r^{{\frac 32}}}$ as $r\to\infty$. The extrinsic curvature for an asymptotically flat initial data set must have fall-off faster than $1/{r^{{\frac3 2}+\epsilon }}$ for  $\epsilon >0$. Asymptotic flatness is required by the Schoen and Yau
proof; the original proof  \cite{Schon:1979rg, Schon:1981vd} actually requires a with stronger fall-off condition on the extrinsic curvature but the result still holds under the weaker fall-off \cite{O'Murchadha:1986}.  $K_{ab}$ also fails an integrability condition used in Witten's proof of the theorem
\cite{Chrusciel:1986bx}. Precisely, this proof shows that the ADM 4-momentum of the spacetime is causal. Demonstrating this  requires that the integral of $K_{ab}K^{ab}$ must be convergent;  the needed fall-off for this is the same as that used in the definition of an asymptotically flat initial data set.  
\end{exam}

One can also construct a similar initial data set which asymptotically approaches a constant mean curvature slice in Schwarzschild spacetime. 
\begin{exam} (An asymptotically constant mean curvature slice)
Again take the spherically symmetric initial data to be of the form (\ref{ssinitial}). Let
\begin{equation}
\xi^{-1} = 1-\gamma(r)\left(\frac{2M_1}{r} -A^2 r^2\right)
\end{equation}
where $\gamma $ is a 
smooth monotonically increasing function which is equal to zero for $r<r_0$ and one for $r>r_1$, 
where the constants $r_0$, $r_1$  are chosen so  that $r_0<r_1$  and $r_0$ is greater than the positive root of
$1-\left(\frac{2M_1}{r} -A^2 r^2\right)$. Note, one can choose $M_1$ to be either positive or negative. Then the constraints (\ref{rhoeqn}) for the vacuum case are satisfied if
\begin{align} 
\beta^2 &= -\frac{\gamma(r)}{r^2}\left(\frac{2M_1}{r} -A^2 r^2\right)\\
\alpha &= \left(1-\gamma(r)\left(\frac{2M_1}{r} -A^2 r^2\right)\right)^{-1}\frac{d\ }{dr} \sqrt{\gamma(r)\left(-\frac{2M_1}{r} +A^2 r^2\right)}
\end{align}
for $r<r_0$, the metric $h_{ab}$ is flat and $K_{ab}=0$. For $r>r_1$, the metric is that of the constant mean curvature  umbilical slice \cite{brill}
${\bf h} = (1-2M_1/r+A^2r^2)^{-1} dr^2 + r^2 d\Omega_2^2$ and the extrinsic curvature approaches
$K_{ab}= Ah_{ab}$. 
Again, it is easy to show, using Birkhoff's theorem, that this spacetime is in fact Minkowski spacetime. 
\end{exam}

\section{Conclusions} 

The local nature of Birkhoff's theorem is important not only for the construction of spacetimes of different global topology but also in identifying  spacetimes arising from the evolution of spherically symmetric initial data sets. Differences  between the local and global characterization of  spherically symmetric spacetimes satisfying Birkhoff's theorem also may result in different physical properties. Birkhoff's theorem  does not imply that a spherically symmetric spacetime has a locally static region (Example \ref{sc}) or is asymptotically flat (Examples  \ref{rp2} and \ref{sc}). Furthermore, not every maximally extended asymptotically flat, spherically symmetric spacetime is the maximal analytic extension of Schwarzschild spacetime (Example \ref{rp3}).
Of course, the universal covering space of these examples, when maximally extended, is the maximal analytic extension of Schwarzchild itself and does have a static region and is asymptotically flat. However, the analogous properties are  lost for the case of positive cosmological constant (Example \ref{sdsc});  the maximal extension of its universal cover  can yield the maximal analytic extension of extremal or overextremal Schwarzschild-de Sitter spacetime, depending on the values of $\Lambda$ and $M$.  These spacetimes have no static region at all and do not exhibit both a future and past asymptotically de Sitter region.
Finally, initial data sets for designer de Sitter spacetimes (Example \ref{designer}), though of  simple geometric form, have arbitrarily complicated topology and their universal covering spacetime need not be de Sitter itself. Hence Birkhoff's theorem allows for a rich variety of locally spherically symmetric spacetimes exhibiting a range of physical properties that do not correspond to all properties of  the global solutions.

Although Birkhoff's theorem applies for $\Lambda<0$, the usual initial data formulation of these spacetimes does not hold for this case as Schwarzschild-anti-de Sitter spacetime is not globally hyperbolic. However, it is clear that globally spherically symmetric examples built from identifications on spacelike hypersurfaces, such as the examples of Section \ref{section4}, readily generalize to this case. Analogues of the locally spherically symmetric examples of of Section \ref{section4a} are also known; the topological black holes of \cite{Aminneborg:1996iz} are locally anti-de Sitter spacetimes and clearly satisfy Birkhoff's theorem.

\section*{Appendix}
The isometries of a
space with identifications can be precisely related to those of a covering space. In particular, one can prove the following theorem 
relating the isometries of a riemannian manifold to those of its universal riemannian covering manifold \cite{Witt:1986ef}.
First, recall that the {\it normalizer} $N_G(H)$ of a group $H\subset G$ is defined as
\begin{equation}N_G(H)=\{ g\in G| gHg^{-1}=H\}\ .\end{equation}
Then
\begin{thm}\label{iso} Given the universal riemannian manifold covering manifold ${\tilde \Sigma}$ of a riemannian manifold
 ${\Sigma}$ with fundamental group $G$, then the isometry groups $Isom({\tilde \Sigma})$ and $Isom({\Sigma})$ 
 are related by the  following:
\begin{equation*}Isom(\Sigma )=\frac{N_{Isom({\tilde \Sigma})}(G)}{ G}\end{equation*}
where N is the normalizer of $G$ in $Isom({\tilde \Sigma})$. 
\end{thm} 
We now use this theorem to compute the isometry group of ${\mathbb R}P^2=S^2/{\mathbb Z}_2$. The isometry group  of $S^2$, $Isom({S^2})$, with the 
round metric is $O(3)$. Now, $O(3)=SO(3)\times {\mathbb Z}_2$ where the ${\mathbb Z}_2$ is generated by
the parity map $P$ on $x\in {\mathbb R}^3$ with $x\rightarrow -x$ via $P$. In fact, this ${\mathbb Z}_2$ action is precisely that which 
produces ${\mathbb R}P^2$ as seen in the first construction.  Next, as ${\mathbb Z}_2=\{ I, P\}$  and as both the identity element and $P$ commute with every element in $O(3)$; $N_{O(3)}({\mathbb Z}_2)=O(3)=SO(3)\times {\mathbb Z}_2$. Hence,
\begin{equation*}Isom({\mathbb R}P^2)=\frac{N_{Isom(S^2)}({\mathbb Z}_2)}{ {\mathbb Z}_2}=
\frac{{N_{O(3)}({\mathbb Z}_2)}}{{\mathbb Z}_2}=\frac{O(3)}{ {\mathbb Z}_2}=\frac{{SO(3)\times {\mathbb Z}_2}}{ {\mathbb Z}_2}
= SO(3)\ .\end{equation*}
Thus ${\mathbb R}P^2$ with round metric is invariant under $SO(3)$; one has not lost continuous isometries, i.e. those generated by the Killing vectors, but only  the discrete symmetry, parity, from its isometry group. 

Next we will use Theorem \ref{iso} to show that the isometry group of ${\mathbb RP}^3 - \{p\}$ includes $SO(3)$.

First, we find the isometry group of ${\mathbb R}P^3 = S^3/{\mathbb Z}_2$.
The isometry group of $S^3$ with the round metric is 
$Isom(S^3)=O(4)$. Thus to apply Theorem \ref{iso}, one must find $N_{O(4)}({\mathbb Z}_2)$ with ${\mathbb Z}_2=\{ I, {\hat P}\}$ as defined in Section \ref{section4}. This is easy in this case; $ N_{O(4)}({\mathbb Z}_2)=O(4)$ as
 ${\hat P}$ commutes with every element in $O(4)$. Then
\begin{equation}
Isom({\mathbb R}P^3)=
\frac {N_{O(4)}({\mathbb Z}_2)} {{\mathbb Z}_2}=\frac{O(4)}{{\mathbb Z}_2}\ .
\end{equation}
This quotient group is more difficult to calculate than for the previous example as $O(4)\neq SO(4)\times {\mathbb Z}_2$.
 However, as ${\mathbb R}P^3$ is orientable, its group of orientation preserving isometries, $Isom^+({\mathbb R}P^3)$, is also related to that of $S^3$ by Theorem \ref{iso}. Furthermore, the full isometry group is the semidirect product of the orientation preserving isometries with a discrete orientation reversion reflection $A$,  $Isom({\mathbb R}P^3) =  Isom^+({\mathbb R}P^3) \ltimes A$. Now, $Isom^+(S^3)=SO(4)$, so
\begin{equation}Isom^+({\mathbb R}P^3)=
\frac {N_{SO(4)}({\mathbb Z}_2)}{ {\mathbb Z}_2}=\frac{SO(4)}{{\mathbb Z}_2}= SO(3)\times SO(3)\ .\end{equation}
As $SO(4)= SU(2)\times SU(2)/ {\mathbb Z}_2$, one sees that, as for the case of ${\mathbb R}P^2$,  only discrete symmetry is lost by the identification.

Next, note that
 $Isom^+( {\mathbb R}P^3 - \{p \})=Isom^+_p({\mathbb R}P^3)$ where $Isom^+_p({\mathbb R}P^3)$
are the isometries which fix the point $p$ on ${\mathbb R}P^3$. Also note that ${\mathbb R}P^3$ is the group manifold $SO(3)$; this follows from the fact that $S^3=SU(3)$ and that ${\mathbb Z}_2=\{ I, {\hat P}\}$
is subgroup of $SU(2)$.
Using this, the explicit action of $SO(3)\times SO(3)$
as an isometry group on ${\mathbb R}P^3$ can be realized by left and right multiplication by elements of $SO(3)$ on  ${\mathbb R}P^3$ realized as the group manifold $SO(3)$; given
$x\in {\mathbb R}P^3$ the isometry acts by $x\to g_1xg_2^{-1}$ for element $(g_1,g_2)$ in $SO(3)\times SO(3)$. Now, without loss of generality, pick the point $p$ to be the identity element $1$ of $SO(3)$. It follows that $Isom^+_p({\mathbb R}P^3)$ will consist of elements $(g_1,g_2)$ that leave the point $p$ fixed; i.e. $g_11 g_2^{-1}=1$. Hence, these are elements of $SO(3)\times SO(3)$ with $g_1=g_2$. Clearly this subgroup is  the group $SO(3)$. Therefore, $Isom^+_p({\mathbb R}P^3)=Isom^+( {\mathbb R}P^3 - \{p \})=SO(3)$.

Finally, we show that the isometry group of the non-orientable handle $S^1 \tilde{ \times} S^2$ contains $SO(3)$. First, note that the non-orientable handle is constructed from $I\times S^2$ by identifying each point on the 2-sphere at one end with its antipodal point on the 2-sphere at the other: alternately $S^1 \tilde{ \times} S^2=S^1\times S^2 /{\mathbb Z}_2$ where the ${\mathbb Z}_2$ is generated by the map  $A:(\psi, S^2) \to (R\psi, P S^2)$ where$R$ is the map given by a rotation by $\pi$ on $S^1$ and  $P$ is the antipodal map on $S^2$. 
Now,  $Isom(S^1 \times S^2)= O(2)\times O(3)$. Furthermore, as $Ag = gA$ for all $g\in O(2)\times O(3)$, the normalizer is 
$ N_{Isom(S^1\times S^2)}({\mathbb Z}_2)=O(2)\times O(3)$. Thus
\begin{equation}
Isom(S^1 \tilde{ \times} S^2)=
\frac {N_{Isom(S^1\times S^2)}({\mathbb Z}_2)}{{\mathbb Z}_2} =\frac{O(2)\times O(4)}{{\mathbb Z}_2}\ .
\end{equation}
This  isometry group clearly contains $SO(3)$.

 Additional explicit examples of using theorem \ref{iso} to calculate isometry groups are given in \cite{Witt:1986ef}.

\acknowledgments The work was supported by NSERC. In addition, the authors would like to
thank the Perimeter Institute for its hospitality.

\end{document}